\documentclass[aps,showpacs,twocolumn,preprintnumbers,amsmath,amssymb]{revtex4}

\UseRawInputEncoding
\usepackage{balance}
\usepackage{hyperref}
\hypersetup{colorlinks=true,
            linkcolor=magenta,
            filecolor=black,
            urlcolor=magenta,
            citecolor=magenta
           }

\usepackage{dcolumn}

\usepackage{float}
\usepackage{graphicx}
\usepackage{epstopdf}
\usepackage{graphicx}
\usepackage{epstopdf}
\usepackage{color}
\usepackage{amsmath,amssymb,amsfonts}

\makeatletter
\newcommand{\rmnum}[1]{\romannumeral #1}
\newcommand{\Rmnum}[1]{\expandafter\@slowromancap\romannumeral #1@}
\makeatother

\begin{document}
\baselineskip=0.5 cm

\title[Article Title]{Effects of tidal charge on Blandford\textbf{\hspace{1pt}\textendash{}\hspace{1pt}}Znajek process around braneworld black holes}

\author{Ruixin Yang$^{1}$, Songbai Chen$^{1,2}$\footnote{Corresponding author: csb3752@hunnu.edu.cn}, Jiliang Jing$^{1,2}$\footnote{jljing@hunnu.edu.cn}}

\affiliation{$^1$Department of Physics, Institute of Interdisciplinary Studies, Hunan Research Center of the Basic Discipline for Quantum Effects and Quantum Technologies, Key Laboratory of Low Dimensional Quantum Structures and Quantum Control of Ministry of Education, Synergetic Innovation Center for Quantum Effects and Applications, Hunan Normal University,  Changsha, Hunan 410081, People's Republic of China
\\ $ ^2$Center for Gravitation and Cosmology, College of Physical Science and Technology, Yangzhou University, Yangzhou 225009, People's Republic of China}
%%%%%%%%%%%%%%%%%%%%%%%%%%%%%%%%%%%%%%%%%%%%%%%%%%%%%%%%%%%%%%%%%%%%%%%%%%%%
\begin{abstract}
\baselineskip=0.4 cm
The Blandford-Znajek (BZ) process is a pivotal mechanism to efficiently extract the energy from a rotating black hole (BH) via its plasma-filled magnetosphere in relativistic astrophysics. Within the framework of extended BZ monopole expansion, we have studied BZ process in the Randall-Sundrum braneworld BH spacetime  and analyzed effects of the tidal charge on the energy and angular momentum extraction rates. It is found that the positive tidal charge reduces the BZ power of a braneworld BH, while the negative tidal charge  enhances the power. Compared with a Kerr BH of the same mass and angular velocity, the BZ power exhibits a maximum reduction of approximately $15.2\%$ in positive cases, whereas in negative cases, it achieves a maximum enhancement of $66.5\%$ in power output. A similar qualitative trend is also observed for the relative angular momentum extraction rate, albeit with different magnitudes.
\end{abstract}
%%%%%%%%%%%%%%%%%%%%%%%%%%%%%%%%%%%%%%%%%%%%%%%%%%%%%%%%%%%%%%%%%%%%%%%%%%%%
\maketitle
%%%%%%%%%%%%%%%%%%%%%%%%%%%%%%%%%%%%%%%%%%%%%%%%%%%%%%%%%%%%%%%%%%%%%%%%%%%%
\section{Introduction}\label{sec:intro}
When a rotating BH is immersed in a magnetic field, Blandford and Znajek demonstrated in their seminal work \cite{Blandford:1977ds} that the energy and angular momentum can be electromagnetically extracted from the BH. Currently, the BZ mechanism stands as the leading theoretical model for explaining relativistic jets launched by supermassive BHs at the center of galaxies, with supporting evidence from the Event Horizon Telescope observations of M87* \cite{EventHorizonTelescope:2019pgp,EventHorizonTelescope:2021srq}.

In general, one can determine the BZ power for a given rotating BH via a regular solution of the force-free Maxwell equations in the BH spacetime. By means of perturbative techniques, the extraction rate was initially established to exhibit a quadratic scaling with the BH angular velocity, $\Omega_{\text{H}}$. However, both analytical \cite{Camilloni:2022kmx} and numerical \cite{Tchekhovskoy:2009ba} studies suggest that the full BZ power expression requires the inclusion of a high-spin factor, typically of the form $f(\Omega_{\text{H}})=1+a\Omega_{\text{H}}^{2}+b\Omega_{\text{H}}^{4}+\cdots$. For a rapidly rotating BH, the higher-order correction terms become relevant. So far, the formula of $f(\Omega_{\text{H}})$ for a monopolar magnetosphere in Kerr spacetime has been extended to the sixth order at most \cite{Camilloni:2022kmx}, and the resulting analytic BZ power expression shows good agreement with both general-relativistic particle-in-cell and magnetohydrodynamic simulations over the full spin range \cite{Meringolo:2025bdu}. It is also noteworthy that the high-spin factor is actually independent of the magnetospheric geometry \cite{Tchekhovskoy:2009ba}. Recently, Camilloni et al. \cite{Camilloni:2023wyn} investigated the BZ jets in the Kerr-MOG spacetime, a rotating BH solution in scalar-tensor-vector gravity, and explicitly derived its $f(\Omega_{\text{H}})$ up to orders $\mathcal{O}(\Omega_{\text{H}}^{2})$, which reveals that those expansion coefficients ($a,b,\dots$) are theory-dependent. To be specific, the high-spin factor of a Kerr-MOG BH is always smaller than that of a Kerr BH with the same mass and angular velocity. More importantly, breaking the parameter degeneracy in $\Omega_{\text{H}}$ enables the function $f(\Omega_{\text{H}})$ to serve as a new probe for testing possible deviations from general relativity (GR) in the strong-field regime \cite{Camilloni:2023wyn}. This motivates us to explore the BZ power in other non-Kerr cases.

We shall focus our attention on a rotating BH in the Randall-Sundrum braneworld scenario, wherein our universe is conceived as a 3-brane embedded in a warped five-dimensional anti-de Sitter bulk \cite{Randall:1999vf}. Within this framework, Aliev and G\"{u}mr\"{u}k\c{c}\"{u}o\u{g}lu \cite{Aliev:2005bi} solved the effective Einstein field equations on the brane, thereby obtaining a stationary and axisymmetric solution of Kerr-Newman type with the usual $Q^{2}$ term replaced by a so-called tidal charge $b$. Consequently, the sign of the tidal charge has a great impact on the global structure of black holes, the orbital dynamics of test particles, and the observational features of such systems \cite{Pun:2008ua,Schee:2008kz,Stuchlik:2008fy,Abdujabbarov:2009az,Aliev:2009cg,Amarilla:2011fx,Aliev:2013jqz,Blaschke:2016uyo,Stuchlik:2017rir,Vagnozzi:2019apd,Khan:2019gco,Banerjee:2019nnj,Nucamendi:2019qsn,Neves:2020doc,Hou:2021okc,Banerjee:2021aln,Du:2021czy,Wei:2022jbi,Banerjee:2022jog,Zi:2024dpi,Liu:2024brf,Kumar:2025njz,Wang:2025fuw}. It is argued that a negative value of $b$ is physically more natural choice because it represents an imprint of the gravitational effects from the bulk space with a negative cosmological constant \cite{Dadhich:2000am,Chamblin:2000ra,Aliev:2005bi}. Moreover, significant research effort has also been devoted to wave-related phenomena in braneworld BH spacetime \cite{deOliveira:2020lzp,deOliveira:2020jha,Dey:2020pth,Chakraborty:2021gdf,Mishra:2021waw,Biswas:2021gvq,Bohra:2023vls}. In \cite{Wei:2022jbi}, Wei et al. compared the influences of tidal charge on energy extraction through magnetic reconnection and BZ mechanism. Nevertheless, the equation employed to calculate BZ power is derived for Kerr BH in the context of standard GR \cite{Tchekhovskoy:2009ba}, which is not a priori valid for braneworld BH, as mentioned before. The main purpose of this work is to study the BZ process rates for Randall-Sundrum braneworld BH, with explicit consideration of the dependence of the high-spin factor on the tidal charge in the energy and angular momentum fluxes.

The manuscript is organized as follows. In the next section, we briefly review the key features of the rotating braneworld BH. To evaluate the BZ rates, in Sec.~\ref{sec:MonoConfig} an extended BZ perturbative approach is used to develop a monopolar force-free magnetosphere around the BH. In Sec.~\ref{sec:BZrates}, we analyze in detail the effects of tidal charge on the total energy and angular momentum fluxes with a non-trivial high-spin factor. Finally, in Sec.~\ref{sec:ConClu} we conclude.

Unless otherwise stated, we adopt geometrized units for which $c=G=1$ and Heaviside-Lorentz units in electromagnetism.
%%%%%%%%%%%%%%%%%%%%%%%%%%%%%%%%%%%%%%%%%%%%%%%%%%%%%%%%%%%%%%%%%%%%%%%%%%%%
\section{Rotating braneworld black hole}\label{sec:RBBBH}
With a stationary and axisymmetric Kerr-Schild ansatz on the brane, the rotating BH model of interest is obtained by solving the effective gravitational field equations for an empty bulk space \cite{Aliev:2005bi}
\begin{equation}
R_{\mu\nu}=-E_{\mu\nu},
\end{equation}
where $R_{\mu\nu}$ is the Ricci tensor, and
\begin{equation}
E_{\mu\nu}={}^{(5)}C_{ABCD}n^{A}n^{C}e^{B}_{\mu}e^{D}_{\nu}
\end{equation}
is the projected ``electric part'' of the five-dimensional Weyl tensor. In Boyer-Lindquist coordinates, the line element of the rotating braneworld BH takes the form \cite{Aliev:2005bi}
\begin{align}\label{eq:metric}
ds^{2}=&-\frac{\Delta-a^{2}\sin^{2}\theta}{\rho^{2}}dt^{2}+\frac{\rho^{2}}{\Delta}dr^{2}+\rho^{2}d\theta^{2}\notag\\
&+\frac{(r^{2}+a^{2})^{2}-\Delta a^{2}\sin^{2}\theta}{\rho^{2}}\sin^{2}\theta d\varphi^{2}\notag\\
&-2a\frac{r^{2}+a^{2}-\Delta}{\rho^{2}}\sin^{2}\theta dtd\varphi,
\end{align}
with
\begin{align}
\rho^{2}&\equiv r^{2}+a^{2}\cos^{2}\theta,\\
\Delta&\equiv r^{2}-2Mr+a^{2}+b,
\end{align}
where $M$, $a$ and $b$ are BH mass, specific angular momentum and tidal charge, respectively. It should be noticed that the parameter $b$, which has the dimension of length squared, may be either positive or negative, unlike its counterpart $Q^{2}$ in Kerr-Newman metric.
Recently, additional low-energy solutions to the Randall-Sundrum II model have been numerically derived  by virtue of the $AdS_5/CFT_4$ dual theory \cite{Figueras:2011gd,Biggs:2021iqw}. Notably, the rotating BHs \cite{Biggs:2021iqw} exhibit four-dimensional behavior and approach the Kerr metric on the brane in the limit of large BHs, while tends toward a five-dimensional Myers-Perry BH with a single non-zero rotation parameter aligned with the brane in the limit of the small ones. Therefore, the solution (\ref{eq:metric}) in the case of the tidal charge $b=0$ is consistent with the limit of large BHs in \cite{Biggs:2021iqw}.

\begin{figure}[htbp]
\centering
\includegraphics[width=0.475\textwidth]{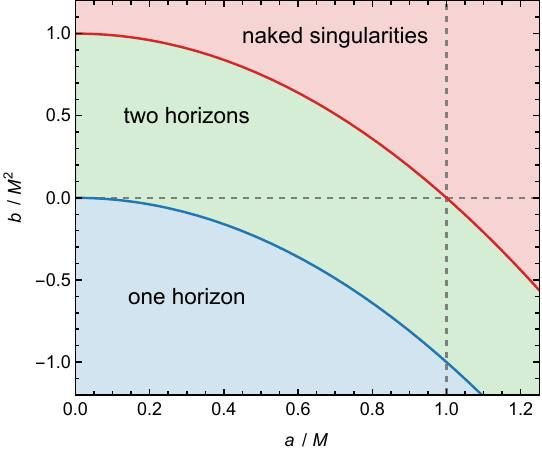}
\caption{The $(a,b)$ plane for braneworld BHs. A key feature is that with a negative $b$, it is possible to set $a>M$ without developing any intrinsic singularities. The region below the horizontal dashed line and to the right of the vertical dashed line corresponds to the GR exclusion zone.}
\label{fig:ParaSpace}
\end{figure}

As expected, the solution \eqref{eq:metric} admits the Killing vectors
\begin{equation}\label{eq:Killing}
(\partial_{t})^{\mu}=(1,0,0,0),\quad(\partial_{\varphi})^{\mu}=(0,0,0,1).
\end{equation}
Two roots of the equation $\Delta=0$ are
\begin{equation}
r_{\pm}=M\pm\sqrt{M^{2}-a^{2}-b}.
\end{equation}
For a fixed $a$, it is evident that a positive/negative $b$ leads to a smaller/larger event horizon radius. In other words, a positive tidal charge reduces the gravity of the BH, whereas a negative one enhances it. This can also be verified by checking the effective Komar mass of the spacetime,
\begin{equation}
M_{\text{eff}}=M-\frac{b}{2r}\left(1+\frac{a^{2}+r^{2}}{ar}\tan^{-1}\frac{a}{r}\right).
\end{equation}
Furthermore, negative tidal charges have the potential to drive a braneworld BH into an over-rotating regime, i.e., the spin $a$ exceeds the mass $M$ (see Fig.~\ref{fig:ParaSpace}), which is never allowed in Einstein theory. In spite of this, a straightforward computation shows that the BH angular velocity
\begin{equation}\label{eq:OmegaH}
\Omega_{\text{H}}=\frac{a}{2Mr_{+}-b}
\end{equation}
is still bounded above by $1/2M$, just as in the Kerr family.

The outer infinite redshift surface is located at
\begin{equation}
r_{\text{e}}=M+\sqrt{M^{2}-a^{2}\cos^{2}\theta-b}.
\end{equation}
Similarly, a negative value of $b$ extends the radius of ergosphere, implying that braneworld BHs with negative tidal charges are supposed to be more energetic objects \cite{Aliev:2005bi}.

%%%%%%%%%%%%%%%%%%%%%%%%%%%%%%%%%%%%%%%%%%%%%%%%%%%%%%%%%%%%%%%%%%%%%%%%%%%%
\section{Monopolar magnetosphere}\label{sec:MonoConfig}
The extraction of rotational energy via BZ mechanism is enabled by the presence of a magnetosphere around a BH. To determine the power associated with such a process, for our purpose it is necessary to solve Maxwell's equations in the background \eqref{eq:metric}. We model the magnetosphere with a monopolar topology here because of the weak dependence of high-spin factor on the specific magnetic field configuration \cite{Tchekhovskoy:2009ba}.
%%%%%%%%%%%%%%%%%%%%%%%%%%%%%%%%%%%%%%%%%%%%%%%%%%%%%%%%%%%%%%%%%%%%%%%%%%%%
\subsection{Force-free electrodynamics}
In the framework of BZ theory \cite{Blandford:1977ds}, a rotating BH is surrounded by a magnetosphere filled with plasma, whose energy density is much lower than that of the electromagnetic field. As a result, the inertia of the plasma itself can be neglected and the following force-free condition approximately holds:
\begin{equation}
F_{\mu\nu}J^{\nu}=0,
\end{equation}
where $F_{\mu\nu}$ is the field intensity 2-form and $J^{\nu}$ the current density of the source. With this constraint, the Maxwell equation $\nabla_{\mu}F^{\mu\nu}=-J^{\nu}$ becomes
\begin{equation}
F_{\mu\nu}\nabla_{\sigma}F^{\sigma\nu}=0,
\end{equation}
or equivalently,
\begin{equation}\label{eq:FFEq}
F_{\mu\nu}\partial_{\sigma}\left(\sqrt{-g}F^{\sigma\nu}\right)=0.
\end{equation}
Another of the Maxwell equations, the Bianchi identities $dF=0$, are automatically satisfied if we choose a gauge potential $A=A_{\mu}dx^{\mu}$ such that $F=dA$. Due to the restriction from the symmetries \eqref{eq:Killing}, all the components $A_{\mu}$ are only functions of $r$ and $\theta$.

To describe a stationary and axisymmetric magnetosphere, in the force-free limit it is convenient to introduce three scalar functions that are invariant along the magnetic field lines \cite{Gralla:2014yja,Camilloni:2022kmx}. The first among them is the magnetic flux (or stream function) passing through a circular loop with radius $r\sin\theta$ centered on the rotational axis of the black hole,
\begin{equation}\label{eq:flux}
\Psi=A_{\varphi}.
\end{equation}
The second one is the angular velocity of the magnetic field lines
\begin{equation}
\Omega=-\frac{\partial_{r}A_{t}}{\partial_{r}A_{\varphi}}=-\frac{\partial_{\theta}A_{t}}{\partial_{\theta}A_{\varphi}}.
\end{equation}
And the third is the poloidal current
\begin{equation}\label{eq:current}
I=\sqrt{-g}F^{\theta r}.
\end{equation}
However, these quantities are not completely independent. Their definitions, together with Eq.~\eqref{eq:FFEq}, yield the integrability conditions \cite{Camilloni:2022kmx}
\begin{align}
(\partial_{r}\Omega)(\partial_{\theta}\Psi)&=(\partial_{\theta}\Omega)(\partial_{r}\Psi),\label{eq:IC1}\\
(\partial_{r}I)(\partial_{\theta}\Psi)&=(\partial_{\theta}I)(\partial_{r}\Psi),\label{eq:IC2}
\end{align}
which imply that the gradients of $\Omega$ and $I$ are everywhere parallel to the gradient of $\Psi$, reducing the functional relations to $\Omega=\Omega(\Psi)$ and $I=I(\Psi)$. In a physical picture, each magnetic field line rigidly rotates with a single angular velocity $\Omega(\Psi)$, while a fixed current $I(\Psi)$ flows along it, both being determined globally by the magnetic flux $\Psi$ the line carries.

Given the above, the general form of a force-free electromagnetic field in the spacetime \eqref{eq:metric} can be expressed as \cite{Gralla:2014yja}
\begin{equation}\label{eq:Faraday}
F=d\Psi\wedge\eta-I\frac{\rho^{2}}{\Delta\sin\theta}dr\wedge d\theta,
\end{equation}
where the 1-form
\begin{equation}
\eta=d\varphi-\Omega dt
\end{equation}
is co-rotating with the magnetic field lines. With the help of \eqref{eq:Faraday}, we may further turn the equations \eqref{eq:FFEq} into \cite{Camilloni:2022kmx}
\begin{align}\label{eq:streameq}
\eta_{\mu}\partial_{r}\left(\eta^{\mu}\Delta\sin\theta\partial_{r}\Psi\right)&+\eta_{\mu}\partial_{\theta}\left(\eta^{\mu}\sin\theta\partial_{\theta}\Psi\right)\notag\\
&+\frac{\rho^{2}}{\Delta\sin\theta}I\frac{dI}{d\Psi}=0,
\end{align}
where $\sqrt{-g}=\rho^{2}\sin\theta$ was used. A physically acceptable solution to the above equation must be regular at the event horizon and asymptotic spacelike infinity, which necessitates the Znajek conditions \cite{Znajek:1977,Komissarov:2004ms,Nathanail:2014aua,Gralla:2014yja,Armas:2020mio,Camilloni:2022kmx}
\begin{align}
I\left(r_{+},\theta\right)&=\left[\frac{r^{2}+a^{2}}{\rho^{2}}\sin\theta\left(\Omega_{\text{H}}-\Omega\right)\partial_{\theta}\Psi\right]\hspace{-1.0pt}\bigg \vert_{\hspace{0pt}r_{+}},\label{eq:ZnajekatEH}\\
I\left(\infty,\theta\right)&=\sin\theta\,\Omega\left(\infty,\theta\right)\partial_{\theta}\Psi\left(\infty,\theta\right).\label{eq:ZnajekatInfty}
\end{align}
In addition, the field lines must smoothly cross the critical surfaces defined by $\eta^{\mu}\eta_{\mu}=0$, namely
\begin{equation}\label{eq:etasquare}
g_{tt}+2\Omega g_{t\varphi}+\Omega^{2}g_{\varphi\varphi}=0.
\end{equation}
As shown in \cite{Komissarov:2004ms}: (\rmnum1) this equation admits only two real, distinct roots, the smaller of which is called the inner light surface (ILS) and the larger of which is called the outer light surface (OLS); (\rmnum2) the ILS is closed and always lies between the event horizon $r_{+}$ and the ergosurface $r_{\text{e}}$, whereas the OLS exhibits a cylindrical topology and does not intersect the ILS at all; (\rmnum3) in the non-rotating limit, the ILS coincides with the event horizon, and the OLS recedes to infinity.
%%%%%%%%%%%%%%%%%%%%%%%%%%%%%%%%%%%%%%%%%%%%%%%%%%%%%%%%%%%%%%%%%%%%%%%%%%%%
\subsection{Static background solutions}
It is difficult to directly solve Maxwell's equations in a general rotating background. However, for small spin $a$, an analytic magnetosphere may be obtained by means of BZ perturbative techniques \cite{Blandford:1977ds}. To this end, we first need a leading-order solution as a seed.

In the static case,  $a=0$, from Eq.~\eqref{eq:metric} it follows that
\begin{equation}\label{eq:RN}
ds^{2}=-\frac{\bar{\Delta}}{r^{2}}dt^{2}+\frac{r^{2}}{\bar{\Delta}}dr^{2}+r^{2}d\theta^{2}+r^{2}\sin^{2}\theta d\varphi^{2},
\end{equation}
where $\bar{\Delta}=(r-\bar{r}_{-})(r-\bar{r}_{+})$, $\bar{r}_{\pm}=M\left(1\pm\sqrt{1-\beta}\right)$, and  $\beta\equiv b/M^{2}$. With this metric, both the angular velocity of the magnetic field lines $\Omega$ and the current $I$ vanish, the stream equation \eqref{eq:streameq} reduces to
\begin{equation}\label{eq:vaceq}
\mathcal{L}\Psi(r,\theta)=0,
\end{equation}
where the self-adjoint operator
\begin{equation}\label{eq:scriptL}
\mathcal{L}=\frac{1}{\sin\theta}\frac{\partial}{\partial r}\left(\frac{\bar{\Delta}}{r^{2}}\frac{\partial}{\partial r}\right)+\frac{1}{r^{2}}\frac{\partial}{\partial\theta}\left(\frac{1}{\sin\theta}\frac{\partial}{\partial\theta}\right).
\end{equation}
For the split-monopole configurations that are symmetric relative to the equatorial plane, the flux must conform to the boundary conditions \cite{Nathanail:2014aua,Armas:2020mio,Camilloni:2022kmx}
\begin{equation}\label{eq:BC}
\Psi(r,0)=1,\quad\Psi(r,\pi/2)=1,\quad\partial_{\theta}\Psi(r,0)=0.
\end{equation}
and must be finite as $r\to\infty$.

Assuming a solution of the form
\begin{equation}
\Psi=c_{0}^{}+\sum_{\ell}R_{\ell}(r)\Theta_{\ell}(\theta),
\end{equation}
Eq.~\eqref{eq:vaceq} can be decoupled into two ordinary differential equations:
\begin{align}
\frac{d}{dr}\left(\frac{\bar{\Delta}}{r^{2}}\frac{dR_{\ell}}{dr}\right)-\frac{\ell(\ell+1)}{r^{2}}R_{\ell}&=0,\label{eq:ridialeq}\\
\frac{d}{d\theta}\left(\frac{1}{\sin\theta}\frac{d\Theta_{\ell}}{d\theta}\right)+\frac{\ell(\ell+1)}{\sin\theta}\Theta_{\ell}&=0.\label{eq:angulareq}
\end{align}
Obviously, the tidal charge $\beta$ remains only in the radial part, while the regular solution to the angular equation \eqref{eq:angulareq} is the same as that in the Schwarzschild spacetime, which is given by \cite{Gralla:2015vta}
\begin{equation}
\Theta_{\ell}=
\begin{cases}
{}_{2}F_{1}\left(\dfrac{\ell}{2},-\dfrac{\ell+1}{2},\dfrac{1}{2};\cos^{2}\theta\right),&\ell\text{ odd}\vspace{5pt}\\
{}_{2}F_{1}\left(-\dfrac{\ell}{2},\dfrac{\ell+1}{2},\dfrac{3}{2};\cos^{2}\theta\right)\cos\theta,&\ell\text{ even}
\end{cases}
,
\end{equation}
where ${}_{2}F_{1}$ denotes the hypergeometric function. For the lowest order ones, e.g.,
\begin{gather}
\Theta_{0}=\cos\theta,\quad\Theta_{1}=\sin^{2}\theta,\\
\Theta_{2}=\cos\theta\sin^{2}\theta,\quad\Theta_{3}=\left(1-5\cos^{2}\theta\right)\sin^{2}\theta.
\end{gather}
On the other hand, Camilloni et al. \cite{Camilloni:2023wyn} reformulated Eq.~\eqref{eq:ridialeq} as a Heun's equation characterized by four regular singular points and showed that its general solution consists in a linear combination of two sets of eigenfunctions,
\begin{equation}\label{eq:ridhar}
R_{\ell}=c_{1}^{}U_{\ell}(w_{\beta};w)+c_{2}^{}V_{\ell}(w_{\beta};w),\quad\ell\geqslant 1,
\end{equation}
where the dimensionless variables
\begin{equation}
w=\frac{r}{\bar{r}_{+}},\quad w_{\beta}=\frac{\bar{r}_{-}}{\bar{r}_{+}}.
\end{equation}
The explicit expressions for $U_{\ell}$ and $V_{\ell}$ can be easily found in Ref.~\cite{Camilloni:2023wyn}, and their asymptotic behaviors at the static event horizon and at spatial infinity read
\begin{align}
U_{\ell}(w_{\beta};1)&\sim(2M)^{\ell+1}(1-w_{\beta})^{\ell},\label{eq:UatEH}\\
V_{\ell}(w_{\beta};1)&\sim(2M)^{-\ell}\log\left(\frac{w-1}{w-w_{\beta}}\right),\\
U_{\ell}(w_{\beta};\infty)&\sim(2Mw)^{\ell+1},\\
V_{\ell}(w_{\beta};\infty)&\sim(2Mw)^{-\ell}.\label{eq:VatInfty}
\end{align}
The radial harmonics \eqref{eq:ridhar} are valid for any Reissner-Nordstr\"{o}m type metric like \eqref{eq:RN}. Collecting these results above, it is clear that the homogeneous equation \eqref{eq:vaceq} does not have a general solution that is regular over the entire space.
%%%%%%%%%%%%%%%%%%%%%%%%%%%%%%%%%%%%%%%%%%%%%%%%%%%%%%%%%%%%%%%%%%%%%%%%%%%%
\subsection{Perturbed configurations}
The extended BZ method allows the force-free field variables to be expanded as a power series in the BH spin parameter $\alpha\equiv a/M$ \cite{Armas:2020mio,Camilloni:2022kmx}:
\begin{align}
\Psi&=\sum_{n=0}^{\infty}\alpha^{n}\Psi_{n}(r,\theta),\label{eq:perPsi}\\
\Omega&=\frac{1}{2M}\sum_{n=0}^{\infty}\alpha^{n}\Omega_{n}(r,\theta),\label{eq:perOmega}\\
I&=\frac{1}{2M}\sum_{n=0}^{\infty}\alpha^{n}I_{n}(r,\theta).\label{eq:perI}
\end{align}
The factor $2M$ is included so that $\Omega_{n}$ and $I_{n}$ are dimensionless. Henceforth, we follow Ref.~\cite{Camilloni:2023wyn} and retain terms up to order $\alpha^{3}$.

As a seed solution, our starting point is the split-monopole configuration
\begin{equation}\label{eq:iniconf}
\Psi_{0}=1-\cos\theta,\quad\Omega_{0}=I_{0}=0,
\end{equation}
where the polar angle ranges between $[0,\pi/2]$. It is easy to verify that $\Psi_{0}$, which is consistent with the boundary conditions \eqref{eq:BC}, is indeed a vacuum solution to Eq.~\eqref{eq:vaceq}.

At the first order in $\alpha$, the integrability conditions \eqref{eq:IC1} and \eqref{eq:IC2} indicate that neither $\Omega_{1}$ nor $I_{1}$ is a function of $r$, meaning
\begin{equation}
\Omega_{1}=\omega_{1}(\theta),\quad I_{1}=i_{1}(\theta).
\end{equation}
Substituting Eqs.~\eqref{eq:perPsi}, \eqref{eq:perOmega} and \eqref{eq:perI} into \eqref{eq:streameq} and keeping only up to the first order terms in the perturbations, we find
\begin{equation}
\mathcal{L}\Psi_{1}(r,\theta)=0,
\end{equation}
where the operator $\mathcal{L}$ is defined by Eq.~\eqref{eq:scriptL}. By virtue of the asymptotic behaviors \eqref{eq:UatEH}--\eqref{eq:VatInfty}, as was mentioned previously, regularity at both the horizon and infinity forces the solution to be trivial
\begin{equation}
\Psi_{1}=0.
\end{equation}
Then, the Znajek conditions \eqref{eq:ZnajekatEH} and \eqref{eq:ZnajekatInfty} at the first order in $\alpha$ give
\begin{equation}
\omega_{1}=\frac{1}{\left(1+\sqrt{1-\beta}\right)^{2}},\quad i_{1}=\omega_{1}\Theta_{1}.
\end{equation}

\begin{figure}[htbp]
\centering
\includegraphics[width=0.475\textwidth]{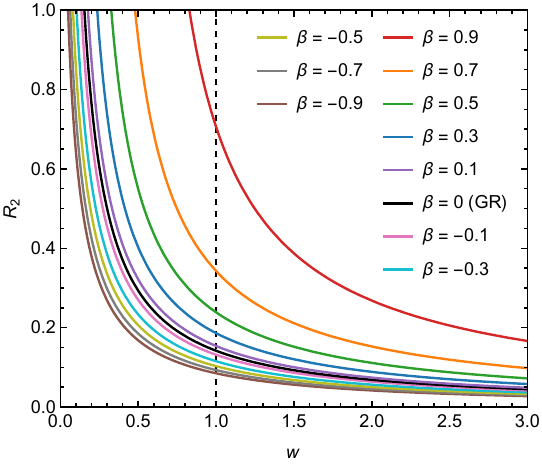}
\caption{$R_{2}$ versus the dimensionless radial coordinate $w$ for different values of $\beta$. All of the curves seamlessly traverse the static event horizon $w=1$ and run towards zero as $w\to\infty$.}
\label{fig:R2}
\end{figure}

Moving to the next perturbative order, it turns out that $\Omega_{2}=\omega_{2}(\theta)$, $I_{2}=i_{2}(\theta)$, and $\Psi_{2}$ obeys the sourced equation
\begin{equation}
\mathcal{L}\Psi_{2}=-\frac{4M^{3}}{r^{4}\bar{r}_{+}^{2}}\left(r-\frac{\bar{r}_{+}\bar{r}_{-}}{2M}\right)\frac{r+\bar{r}_{+}}{r-\bar{r}_{-}}\frac{\Theta_{2}}{\sin\theta}.
\end{equation}
A particular solution to this equation is found to be $\Psi_{2}=R_{2}(r)\Theta_{2}(\theta)$ with \cite{Camilloni:2023wyn}
\begin{align}
R_{2}&=-\frac{1}{2w_{\beta}}\frac{\left(1+w_{\beta}\right)^{2}}{\left(1-w_{\beta}\right)^{2}}\left\{2w^{2}\left(1+3w_{\beta}\right)\phantom{\left[\operatorname{Li}_{2}\left(\frac{1-w}{1-w_{\beta}}\right)\right]}\right.\notag\\
&-\frac{w\left(1+w_{\beta}\right)\left(3+w_{\beta}\right)}{2}-\frac{w_{\beta}\left[17+w_{\beta}\left(20-w_{\beta}\right)\right]}{9}\notag\\
&-\frac{6w\left[4w-\left(1+w_{\beta}\right)\right]-\left[1+w_{\beta}\left(10+w_{\beta}\right)\right]}{6}\notag\\
&\times\left[\log\left(w\right)-\left(1+w_{\beta}\right)\log\left(\frac{w-w_{\beta}}{1-w_{\beta}}\right)\right]\notag\\
&-\frac{4w^{3}-\left(1+w_{\beta}\right)\left(3w^{2}-w_{\beta}\right)}{1-w_{\beta}}\left[\operatorname{Li}_{2}\left(\frac{w_{\beta}}{w}\right)\phantom{\log\left(\frac{w-1}{w-w_{\beta}}\right)}\right.\notag\\
&-\operatorname{Li}_{2}\left(\frac{1}{w}\right)+\log\left(w\right)\log\left(\frac{w-1}{w-w_{\beta}}\right)\notag\\
&+\left(1+w_{\beta}\right)\left(\frac{\pi^{2}}{6}+\frac{1}{2}\log^{2}\left(\frac{w-w_{\beta}}{1-w_{\beta}}\right)-\frac{1-w_{\beta}}{2w_{\beta}}\right.\notag\\
&\times\left.\left.\left.\!\!\log\left(1-\frac{w_{\beta}}{w}\right)+\operatorname{Li}_{2}\left(\frac{1-w}{1-w_{\beta}}\right)\right)\right]\right\},\label{eq:R2}
\end{align}
where $\operatorname{Li}_{2}$ denotes the dilogarithm function. Recall that parameter $w_{\beta}$ is linked to the tidal charge via $w_{\beta}=(1-\sqrt{1-\beta})/(1+\sqrt{1-\beta})$, in the limit of $w_{\beta}\to 0$, Eq.~\eqref{eq:R2} reduces to its corresponding form derived for the Kerr geometry \cite{Blandford:1977ds,Camilloni:2022kmx},
\begin{align}
R_{2}={}&\frac{11}{72}+\frac{1}{6w}+w\left(1-2w\right)+\frac{1+6w-24w^{2}}{12}\notag\\
&\times\log\left(w\right)+\frac{w^{2}\left(4w-3\right)}{2}\left[\operatorname{Li}_{2}\left(\frac{1}{w}\right)\right.\notag\\
&\left.-\log\left(w\right)\log\left(1-\frac{1}{w}\right)\right],\label{eq:R2GR}
\end{align}
as a check. These kinds of solutions are obtained by using the Green's function method. For details, see, e.g., Appendix B in Ref.~\cite{Camilloni:2022kmx}. To visualize the behavior of the function $R_{2}(w)$ we plot it in Fig.~\ref{fig:R2}, which also compares it with the standard GR case \eqref{eq:R2GR} marked by a black solid line.
Further, the Znajek conditions \eqref{eq:ZnajekatEH} and \eqref{eq:ZnajekatInfty} tell us that
\begin{equation}
\omega_{2}=i_{2}=0.
\end{equation}

At the order of $\alpha^{3}$, the integrability conditions \eqref{eq:IC1} and \eqref{eq:IC2} lead to $\Omega_{3}=\omega_{3}(\theta)$, and
\begin{equation}
I_{3}=\frac{\partial_{\theta}i_{1}}{\partial_{\theta}\Psi_{0}}\Psi_{2}(r,\theta)+i_{3}(\theta).
\end{equation}
From the stream equation \eqref{eq:streameq} we get $\mathcal{L}\Psi_{3}(r,\theta)=0$, which implies, once again, that
\begin{equation}
\Psi_{3}=0.
\end{equation}
Finally, the Znajek conditions \eqref{eq:ZnajekatEH} and \eqref{eq:ZnajekatInfty} produce
\begin{align}
\omega_{3}={}&\frac{1}{\left(1+\sqrt{1-\beta}\right)^{4}}\left\{\frac{1}{\sqrt{1-\beta}}+\frac{1}{2}\right.\notag\\
&\times\left.\left[1-\left(1+\sqrt{1-\beta}\right)^{2}R_{2}^{\text{H}}\right]\Theta_{1}\right\},\\
i_{3}={}&\frac{1}{\left(1+\sqrt{1-\beta}\right)^{2}}\left\{\left[\frac{1}{\left(1+\sqrt{1-\beta}\right)^{2}}\right.\right.\notag\\
&\times\left.\left(\frac{2}{5}+\frac{1}{\sqrt{1-\beta}}\right)-\frac{2}{5}R_{2}^{\text{H}}\hspace{-4.8em}\phantom{\frac{1}{\left(\sqrt{1-\beta}\right)^{2}}}\right]\Theta_{1}\notag\\
&+\left.\frac{1}{10}\left[\frac{1}{\left(1+\sqrt{1-\beta}\right)^{2}}-R_{2}^{\text{H}}\right]\Theta_{3}\right\},
\end{align}
where $R_{2}^{\text{H}}$ is defined as the function \eqref{eq:R2} evaluated at the static event horizon $w=1$.

\begin{figure*}[htbp]
\centering
\includegraphics[width=0.475\textwidth]{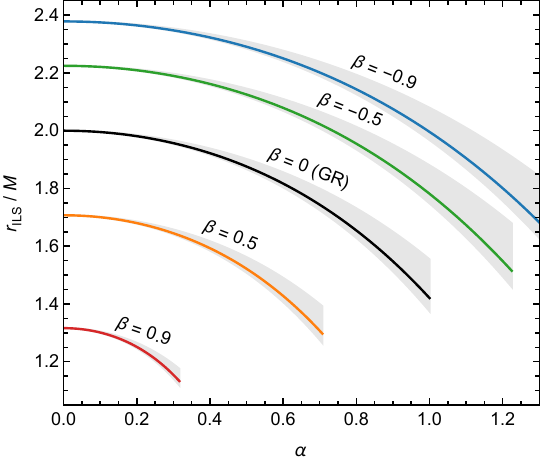}
\includegraphics[width=0.475\textwidth]{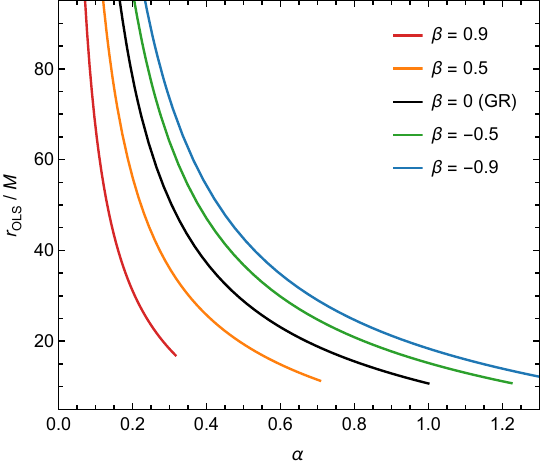}
\caption{Plots of $r_{\text{ILS}}^{}$ and $r_{\text{OLS}}^{}$ as functions of the dimensionless spin parameter $\alpha$ for selected values of $\beta$, with $\theta=30^{\circ}$. The right ends of the lines are cut off because for each $\beta$, there exists a physical bound $\sqrt{1-\beta}$ on $\alpha$, as shown in Fig.~\ref{fig:ParaSpace}. In the left panel, the gray shading delineates the extent of the ergoregion at the same perturbative order as Eq.~\eqref{eq:rILS}.}
\label{fig:lightsurfaces}
\end{figure*}

With several $\Omega_{n}$ in hand, we may now determine the light-surface positions perturbatively by solving the algebraic equation \eqref{eq:etasquare}. For ILS, the result is
\begin{equation}\label{eq:rILS}
\frac{r_{\text{ILS}}^{}}{M}=x_{0}^{}+\alpha^{2}x_{2}^{}+\alpha^{4}x_{4}^{}+\mathcal{O}\left(\alpha^{5}\right),
\end{equation}
in which the coefficients are
\begin{align}
x_{0}^{}&=1+\sqrt{1-\beta},\\
x_{2}^{}&=-\frac{1}{2\sqrt{1-\beta}}\left(1-\frac{\Theta_{1}}{4}\right),\\
x_{4}^{}&=-\frac{1}{8\left(1-\beta\right)^{3/2}}\left\{1\phantom{\frac{8}{1+\sqrt{1-\beta}}}\right.\notag\\
&-\left[1+\frac{8}{1+\sqrt{1-\beta}}+16\left(1-\beta\right)R_{2}^{\text{H}}\right]\frac{\Theta_{1}}{20}\notag\\
&\left.+\left[9-\frac{8}{1+\sqrt{1-\beta}}-16\left(1-\beta\right)R_{2}^{\text{H}}\right]\frac{\Theta_{3}}{80}\right\}.
\end{align}
Similarly for the outer one,
\begin{gather}
\frac{r_{\text{OLS}}^{}}{M}=\frac{y_{-1}^{}}{\alpha}+y_{0}^{}+\alpha y_{1}^{}+\mathcal{O}\left(\alpha^{2}\right),\\
\begin{align}
y_{-1}^{}&=\left(1+\sqrt{1-\beta}\right)^{2}\frac{2}{\sin\theta},\quad y_{0}^{}=-1,\\
y_{1}^{}&=-\left\{\frac{2}{\sqrt{1-\beta}}+\left[\frac{6+8\sqrt{1-\beta}+5\left(1-\beta\right)}{4\left(1+\sqrt{1-\beta}\right)^{2}}\right.\right.\notag\\
&\left.\left.\phantom{\frac{\sqrt{\beta}}{\left(\sqrt{\beta}\right)^{2}}}-\left(1+\sqrt{1-\beta}\right)^{2}R_{2}^{\text{H}}\right]\Theta_{1}\right\}\frac{1}{\sin\theta}.
\end{align}
\end{gather}
The relations between the two light surfaces and the black hole spin are illustrated in Fig.~\ref{fig:lightsurfaces}. For a fixed tidal charge $\beta$, the values of both $r_{\text{ILS}}^{}$ and $r_{\text{OLS}}^{}$ decrease as $\alpha$ increases. What differs, however, is that $r_{\text{ILS}}^{}$ begins its drop at the static horizon $\bar{r}_{+}$, whereas $r_{\text{OLS}}^{}$ is dragged in from infinity to a finite distance by the rotation. For a given $\alpha$, a positive $\beta$ leads to smaller $r_{\text{ILS}}^{}$ and $r_{\text{OLS}}^{}$, while a negative $\beta$ has the opposite effect.

\begin{figure*}[htbp]
\centering
\begin{minipage}[b]{\textwidth}
\centering
\includegraphics[width=0.32\textwidth]{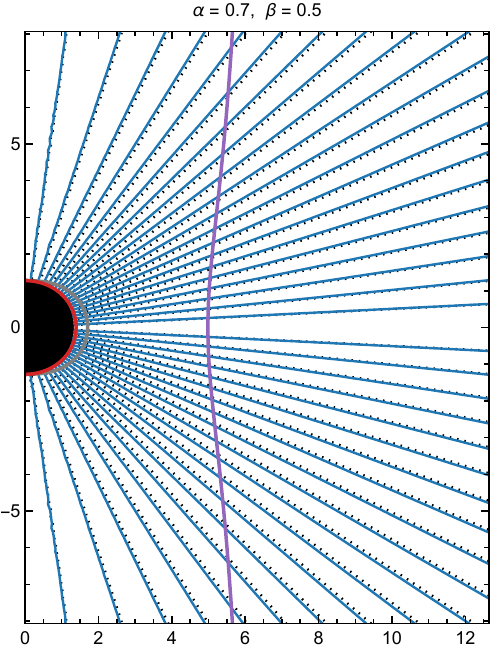}
\includegraphics[width=0.32\textwidth]{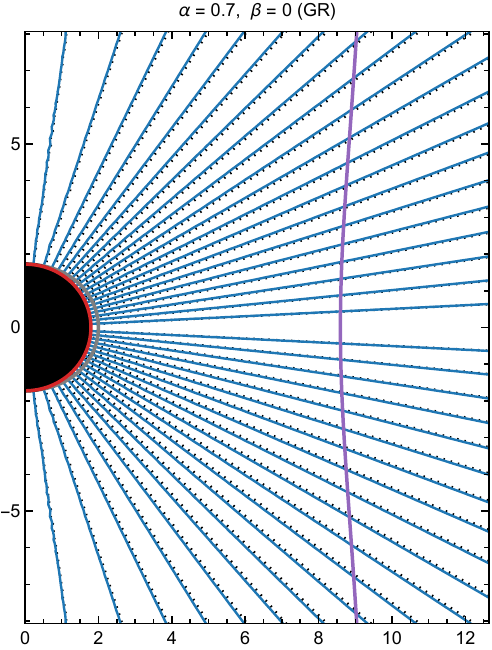}
\includegraphics[width=0.32\textwidth]{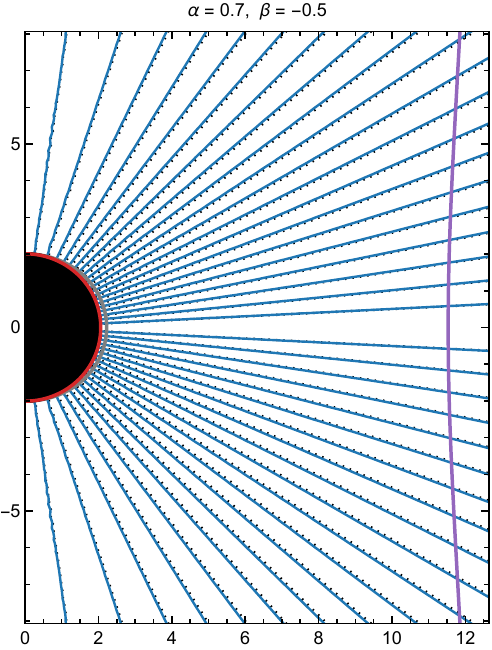}
\end{minipage}
%\hfill
\begin{minipage}[b]{\textwidth}
\centering
\includegraphics[width=0.32\textwidth]{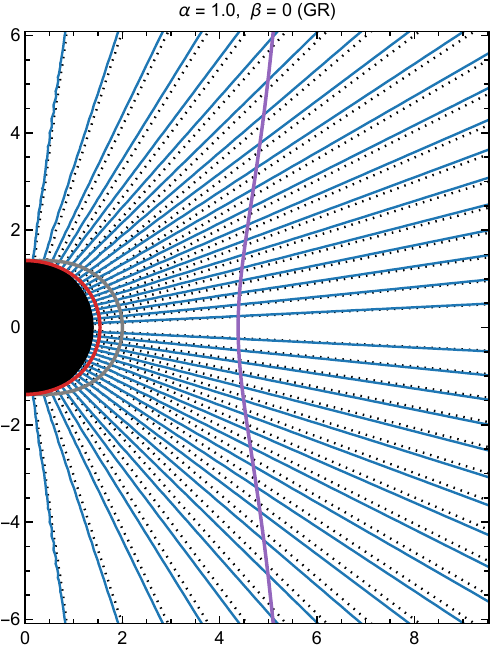}
\includegraphics[width=0.32\textwidth]{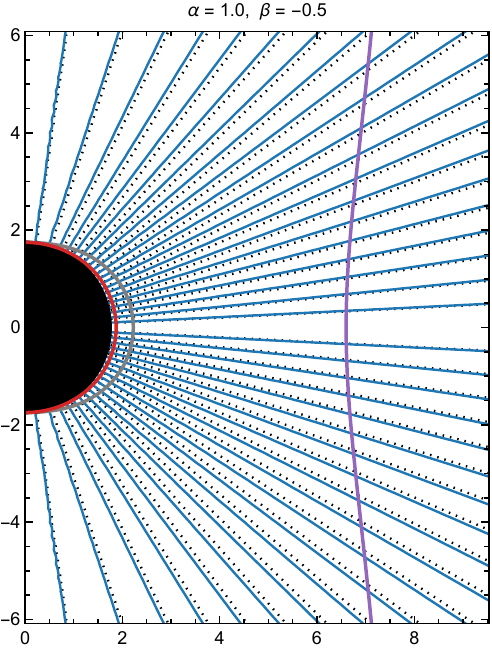}
\includegraphics[width=0.32\textwidth]{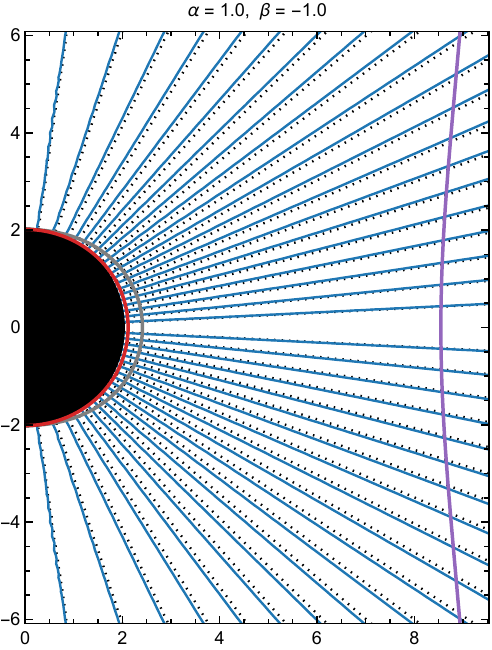}
\end{minipage}
%\hfill
\begin{minipage}[b]{\textwidth}
\centering
\includegraphics[width=0.32\textwidth]{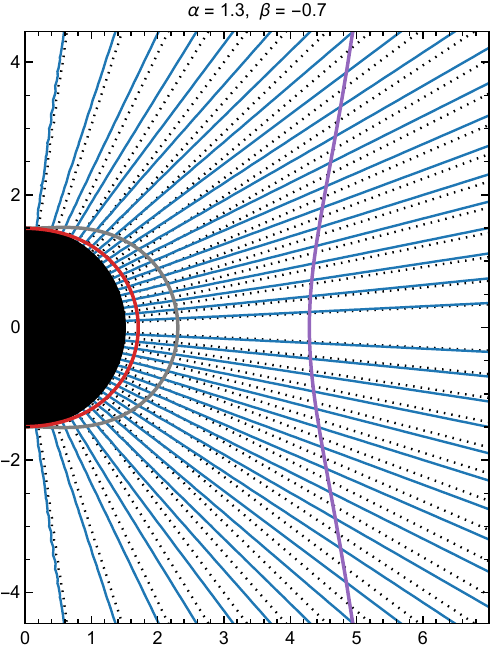}
\includegraphics[width=0.32\textwidth]{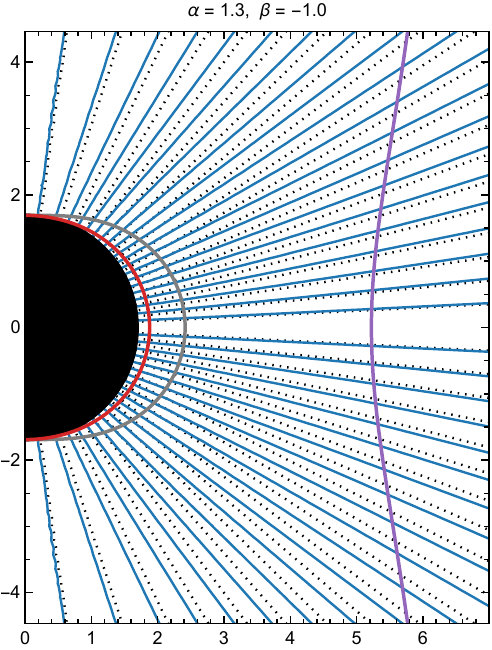}
\includegraphics[width=0.32\textwidth]{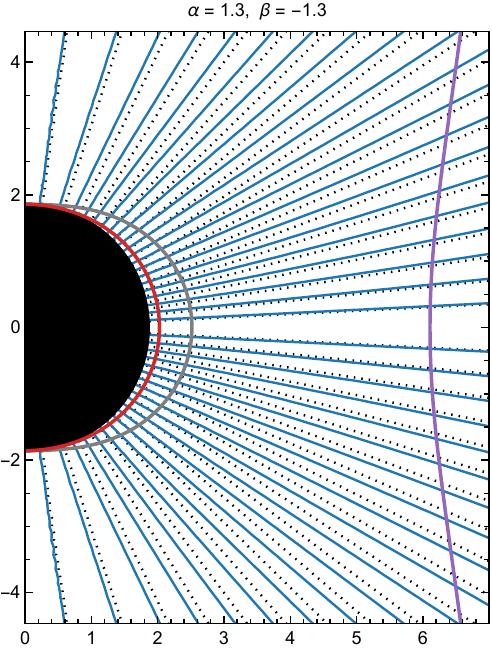}
\end{minipage}
\caption{Magnetic field lines as the contours of flux $\Psi$ on the $(r,\theta)$ plane for $\alpha=0.7$ (top row), $\alpha=1.0$ (middle row), and $\alpha=1.3$ (bottom row). The leading-order solution \eqref{eq:iniconf} is also included and depicted with dotted lines. In addition, the gray, red, and purple curves are used to identify the ergosurface, ILS, and OLS, respectively.}
\label{fig:FieldLines}
\end{figure*}

\begin{figure}[htbp]
\centering
\includegraphics[width=0.475\textwidth]{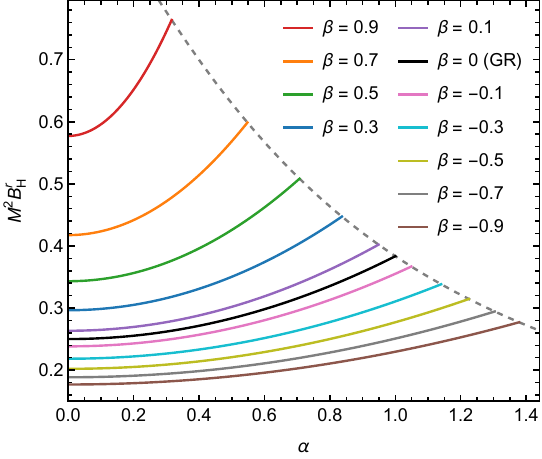}
\caption{The radial magnetic field $B^{r}=\left(\partial_{\theta}\Psi\right)/\sqrt{-g}$ at the north pole of the braneworld BH horizon. A dashed line defines the boundary of physically acceptable configurations.}
\label{fig:BrH}
\end{figure}

Fig.~\ref{fig:FieldLines} shows the obtained monopolar magnetospheres around braneworld BHs in different rotating regimes --- specifically, $\alpha<1$, $\alpha=1$, and $\alpha>1$. Observe that all the magnetic field lines escape from the event horizon, extend to spatial infinity, and cross the critical surfaces in a well-behaved manner. Compared with the initial configuration, the high-order field lines have a tendency to bunch up towards the BH's rotation axis. The larger $\alpha$ or $\beta$ is, the closer they are to the axis, thereby enhancing the radial magnetic fields on the event horizon (see Fig.~\ref{fig:BrH}). In Ref.~\cite{Gralla:2015uta}, the bunching of field lines in BH magnetospheres with various geometries was explored analytically.

%%%%%%%%%%%%%%%%%%%%%%%%%%%%%%%%%%%%%%%%%%%%%%%%%%%%%%%%%%%%%%%%%%%%%%%%%%%%
\section{Blandford\textbf{\hspace{1pt}\textendash{}\hspace{1pt}}Znajek rates}\label{sec:BZrates}
For a stationary and axisymmetric magnetosphere, there are two Noether currents associated with the Killing fields \eqref{eq:Killing}, i.e.,
\begin{equation}
\mathcal{J}^{E}_{\mu}=-T_{\mu\nu}(\partial_{t})^{\nu},\quad\mathcal{J}^{L}_{\mu}=T_{\mu\nu}(\partial_{\varphi})^{\nu},
\end{equation}
where the stress-energy tensor is of course
\begin{equation}
T_{\mu\nu}=F_{\mu\sigma}F_{\nu}{}^{\sigma}-\frac{1}{4}g_{\mu\nu}F_{\sigma\rho}F^{\sigma\rho}.
\end{equation}
Physically, $\mathcal{J}^{E}_{\mu}$ is the energy flux density and $\mathcal{J}^{L}_{\mu}$ the angular momentum flux density. Integrating the dual form of $\mathcal{J}^{E}_{\mu}$ over a hypersurface generated by a sphere $S^{2}$ of radius $r$ cross a lapse $\Delta t$ of time, in the $\Delta t\to 0$ limit, gives the total flux of energy flowing out of that sphere \cite{Gralla:2014yja,Camilloni:2022kmx},
\begin{equation}\label{eq:dotE}
\dot{E}=2\pi\int_{0}^{\pi}\Omega(r,\theta)I(r,\theta)\partial_{\theta}\Psi(r,\theta)d\theta,
\end{equation}
where the overdot stands for the derivatives with respect to the Killing time $t$. Similarly, the outward angular momentum flux reads \cite{Gralla:2014yja,Camilloni:2022kmx}
\begin{equation}\label{eq:dotL}
\dot{L}=2\pi\int_{0}^{\pi}I(r,\theta)\partial_{\theta}\Psi(r,\theta)d\theta.
\end{equation}
The energy and angular momentum extraction rates from a rotating BH via the BZ process are determined by Eqs.~\eqref{eq:dotE} and \eqref{eq:dotL}, respectively. Since the fluxes are conserved, one can evaluate them at the event horizon $r=r_{+}$.

Feeding the invariants \eqref{eq:perPsi}--\eqref{eq:perI} into Eq.~\eqref{eq:dotE}, the power is found to be
\begin{equation}\label{eq:powerofalpha}
\dot{E}=\alpha^{2}\dot{E}_{2}+\alpha^{4}\dot{E}_{4}+\mathcal{O}\left(\alpha^{6}\right),
\end{equation}
with
\begin{align}
\dot{E}_{2}={}&\frac{2\pi}{3M^{2}\left(1+\sqrt{1-\beta}\right)^{4}},\\
\dot{E}_{4}={}&\frac{4\pi}{15M^{2}\left(1+\sqrt{1-\beta}\right)^{6}}\left\{\frac{5}{\sqrt{1-\beta}}\right.\notag\\
&\left.\ \,+\,2\left[1-\left(1+\sqrt{1-\beta}\right)^{2}\!R_{2}^{\text{H}}\right]\right\}.
\end{align}
Along the same lines, from Eq.~\eqref{eq:dotL} we obtain
\begin{align}
\dot{L}&=\alpha\dot{L}_{1}+\alpha^{3}\dot{L}_{3}+\mathcal{O}\left(\alpha^{5}\right),\label{eq:dotLofalpha}\\
\dot{L}_{1}&=2M\left(1+\sqrt{1-\beta}\right)^{2}\dot{E}_{2},\\
\dot{L}_{3}&=M\left(1+\sqrt{1-\beta}\right)^{2}\dot{E}_{4}.
\end{align}
If we set $\beta=0$, Eqs.~\eqref{eq:powerofalpha} and \eqref{eq:dotLofalpha} reduce to the energy and angular momentum fluxes derived in Kerr spacetime \cite{Tanabe:2008wm,Camilloni:2022kmx}, at their respective orders.

To find out the high-spin factor, with the help of Eq.~\eqref{eq:OmegaH} we can recast the power \eqref{eq:powerofalpha} as an expansion in the BH angular frequency. In fact, it has been argued in \cite{Tchekhovskoy:2009ba} that an $\Omega_{\text{H}}$-expansion for the BZ rates is much more natural and accurate than the $\alpha$-expansion was before. Up to the fourth order in $\Omega_{\text{H}}$, we arrive at
\begin{equation}\label{eq:powerofOmegaH}
\dot{E}=\frac{2}{3}\pi\Omega_{\text{H}}^{2}f_{\beta}^{}\left(\Omega_{\text{H}}\right),
\end{equation}
where
\begin{align}\label{eq:HSfactor}
f_{\beta}^{}=1&+\frac{4}{5}M^{2}\Omega_{\text{H}}^{2}\left(1+\sqrt{1-\beta}\right)^{2}\notag\\
&\times\left[1-\left(1+\sqrt{1-\beta}\right)^{2}\!R_{2}^{\text{H}}\right]+\mathcal{O}\left(\Omega_{\text{H}}^{4}\right).
\end{align}
The subscript $\beta$ serves to highlight the background dependence of the high-spin factor, as noted in \cite{Camilloni:2023wyn}. In the $\beta\to 0$ limit,
\begin{equation}
f_{0}^{}=1+\frac{8}{45}\left(67-6\pi^{2}\right)M^{2}\Omega_{\text{H}}^{2}+\mathcal{O}\left(\Omega_{\text{H}}^{4}\right),
\end{equation}
the BZ4 formula given in Ref.~\cite{Tchekhovskoy:2009ba} is recovered by aligning the relevant coefficients carefully. For the angular momentum flux, it follows from Eq.~\eqref{eq:dotLofalpha} that
\begin{equation}\label{eq:dotLofOmegaH}
\dot{L}=\frac{2}{3}\pi\Omega_{\text{H}}\left(1+f_{\beta}^{}\right).
\end{equation}
To the best of our knowledge, Eqs.~\eqref{eq:powerofOmegaH} and \eqref{eq:dotLofOmegaH} are also valid for other neutral, single-parameter modified Kerr BHs with a monopolar magnetosphere. Moreover, in the absence of the high-spin factor, the remaining $\dot{E}=2\pi\Omega_{\text{H}}^{2}/3$ and $\dot{L}=4\pi\Omega_{\text{H}}/3$ are exactly the same as the leading-order results obtained in standard GR \cite{Camilloni:2022kmx}. That is to say, the BZ mechanism cannot distinguish a Kerr BH from a braneworld one under consideration unless the high-spin factor \eqref{eq:HSfactor} is measured, due to the degeneracy of the spin parameter $a$ and the tidal charge $b$ in Eq.~\eqref{eq:OmegaH}. Indeed, the degeneracy in the leading-order term of the BZ power has been confirmed in a general stationary and axisymmetric BH spacetime \cite{Konoplya:2021qll}.

\begin{figure}[htbp]
\centering
\includegraphics[width=0.475\textwidth]{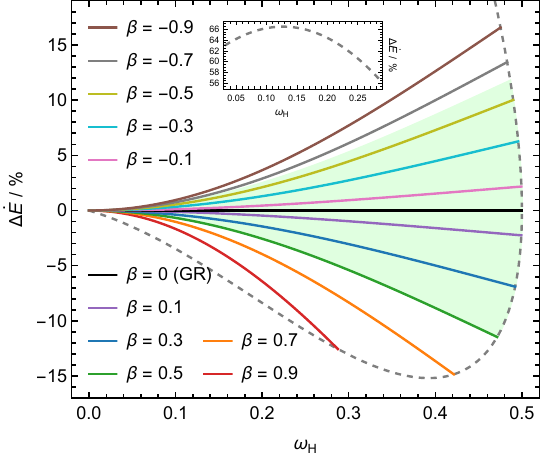}
\caption{Percentage-scaled $\Delta\dot{E}$ as a function of the dimensionless angular velocity $\omega_{\text{H}}\equiv M\Omega_{\text{H}}$ for different values of $\beta$, where the independent variable is bounded by $\sqrt{1-\beta}/(2-\beta)$ according to Eq.~\eqref{eq:OmegaH}. Since there are no allowed configurations outside the dashed gray line, $\Delta\dot{E}$ reaches its minimum of approximately $-15.2\%$ at $\omega_{\text{H}}=0.392$ and $\beta=0.765$. The inset shows the envelope at extremely negative tidal charges, revealing a maximum enhancement $\Delta\dot{E}_{\text{max}}\approx 66.5\%$ for $\omega_{\text{H}}=0.127$ with $\beta=-58.818$. Configurations in the light green area are compatible with the BH shadow observations \cite{Banerjee:2019nnj,Banerjee:2022jog,Krishnendu:2024jkj}}.
\label{fig:DeltadotE}
\end{figure}

\begin{figure}[htbp]
\centering
\includegraphics[width=0.475\textwidth]{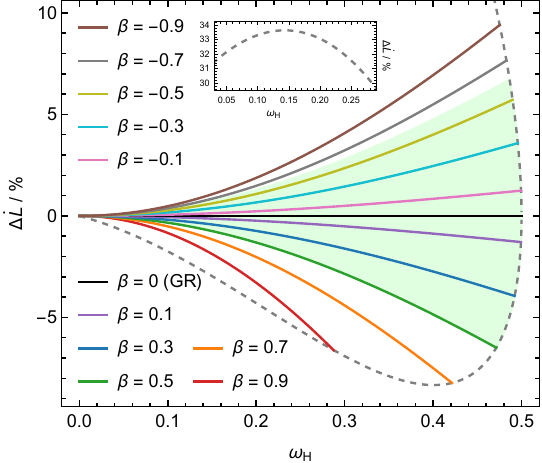}
\caption{Relative difference in angular momentum extraction rate, $\Delta\dot{L}$, between Kerr and braneworld Kerr BHs with identical $M$ and $\Omega_{\text{H}}$. The function \eqref{eq:DeltadotL} exhibits a maximum of $33.7\%$ at $\omega_{\text{H}}=0.143$, $\beta=-45.602$, and a minimum of $-8.3\%$ at $\omega_{\text{H}}=0.401$, $\beta=0.749$. The light green area delineates the observational constraints $\beta\in[-0.6,0.5]$ \cite{Banerjee:2019nnj,Banerjee:2022jog,Krishnendu:2024jkj}}.
\label{fig:DeltadotL}
\end{figure}

What is really of concern, as suggested in Ref.~\cite{Camilloni:2023wyn}, is the fractional deviation of the high-spin factor $f_{\beta}^{}$ with respect to its GR limit $f_{0}^{}$, or equivalently, is the relative difference in BZ power between a Kerr and a braneworld BH with the same mass $M$ and angular velocity $\Omega_{\text{H}}$,
\begin{equation}\label{eq:DeltadotE}
\Delta\dot{E}=\frac{f_{\beta}^{}-f_{0}^{}}{f_{0}^{}}.
\end{equation}
It is clear from Fig.~\ref{fig:DeltadotE} that a positive $\beta$ always leads to a negative $\Delta\dot{E}$, and therefore tends to reduce the energy extraction rate. With a high-spin factor accurate to the second order in $\Omega_{\text{H}}$, the effect of positive tidal charges can suppress the BZ power of a braneworld BH by up to approximately $15.2\%$ compared to that of a Kerr BH. Conversely, a negative tidal charge acts to boost the power by $66.5\%$ at most from a purely theoretical perspective. Such an enhancement, however, can only be achieved at $\beta\approx-58.8$, a value that is likely not allowed in realistic astrophysical contexts. For example, observations of the BH shadow constrain the dimensionless tidal charge to $\beta=-0.1^{+0.6}_{-0.5}$ \cite{Banerjee:2019nnj,Banerjee:2022jog,Krishnendu:2024jkj}, corresponding to a maximum $\Delta\dot{E}$ of $11.8\%$.
And then we finally turn to the relative angular momentum flux,
\begin{equation}\label{eq:DeltadotL}
\Delta\dot{L}=\frac{f_{0}^{}}{1+f_{0}^{}}\Delta\dot{E},
\end{equation}
which behaves qualitatively similarly to Eq.~\eqref{eq:DeltadotE}, as presented in Fig.~\ref{fig:DeltadotL}. Moreover, the light green areas in Figs.~\ref{fig:DeltadotE} and \ref{fig:DeltadotL} correspond to the allowed regions from the BH shadow observations for the relative energy flux $\Delta\dot{E}$ and the relative angular momentum flux $\Delta\dot{L}$, respectively.
%%%%%%%%%%%%%%%%%%%%%%%%%%%%%%%%%%%%%%%%%%%%%%%%%%%%%%%%%%%%%%%%%%%%%%%%%%%%
\section{Conclusions}\label{sec:ConClu}
This work provides an analytical investigation into the effects of tidal charge on the BZ process rates in a rotating braneworld BH spacetime. To do so, following Ref.~\cite{Camilloni:2023wyn}, we expand the variables $\Psi$, $\Omega$, and $I$ as a truncated power series in the dimensionless spin parameter $\alpha$, and thereby obtain a monopolar BH magnetosphere that is regular at all the critical surfaces by solving the force-free Maxwell equations perturbatively. For higher precision, we recast the BZ rates as expansions in the BH angular velocity $\Omega_{\text{H}}$, which yields a high-spin factor manifestly dependent on the tidal charge.

For Kerr and braneworld BHs sharing the same mass $M$ and angular velocity $\Omega_{\text{H}}$, it is found that the BZ power is suppressed by the positive tidal charge, with a maximum reduction of approximately $15.2\%$ given our current level of precision, whereas a negative tidal charge can enhance the power by $66.5\%$ at most theoretically. A similar trend also holds for the angular momentum extraction rate, which is suppressed by a positive tidal charge and enhanced by a negative one, with the extremal values of $-8.3\%$ and $33.7\%$, respectively. These results are helpful in placing independent constraints on the tidal charge parameter and the extra dimension size through future horizon-scale observations.

A direct extension of our work would be to incorporate higher-order contributions to the high-spin factor for the braneworld BH, alongside a comparative study of its BZ process with other known energy extraction mechanisms.
%%%%%%%%%%%%%%%%%%%%%%%%%%%%%%%%%%%%%%%%%%%%%%%%%%%%%%%%%%%%%%%%%%%%%%%%%%%%
\begin{acknowledgments}
This work was  supported by the National Natural Science Foundation of China under Grant No.12275078, 11875026, 12035005, 2020YFC2201400 and the innovative research group of Hunan Province under Grant No. 2024JJ1006.
\end{acknowledgments}
%%%%%%%%%%%%%%%%%%%%%%%%%%%%%%%%%%%%%%%%%%%%%%%%%%%%%%%%%%%%%%%%%%%%%%%%%%%%
\clearpage
%%%%%%%%%%%%%%%%%%%%%%%%%%%%%%%%%%%%%%%%%%%%%%%%%%%%%%%%%%%%%%%%%%%%%%%%%%%%
\bibliography{bz_bbh}

@article{Blandford:1977ds,
    author = "Blandford, R. D. and Znajek, R. L.",
    title = "{Electromagnetic extractions of energy from Kerr black holes}",
    doi = "10.1093/mnras/179.3.433",
    journal = "Mon. Not. Roy. Astron. Soc.",
    volume = "179",
    pages = "433--456",
    year = "1977"
}

@article{EventHorizonTelescope:2019pgp,
    author = "Akiyama, Kazunori and others",
    collaboration = "Event Horizon Telescope",
    title = "{First M87 Event Horizon Telescope Results. V. Physical Origin of the Asymmetric Ring}",
    doi = "10.3847/2041-8213/ab0f43",
    journal = "Astrophys. J. Lett.",
    volume = "875",
    number = "1",
    pages = "L5",
    year = "2019"
}

@article{EventHorizonTelescope:2021srq,
    author = "Akiyama, Kazunori and others",
    collaboration = "Event Horizon Telescope",
    title = "{First M87 Event Horizon Telescope Results. VIII. Magnetic Field Structure near The Event Horizon}",
    reportNumber = "FERMILAB-PUB-21-850-PPD",
    doi = "10.3847/2041-8213/abe4de",
    journal = "Astrophys. J. Lett.",
    volume = "910",
    number = "1",
    pages = "L13",
    year = "2021"
}

@article{Camilloni:2022kmx,
    author = "Camilloni, Filippo and Dias, Oscar J. C. and Grignani, Gianluca and Harmark, Troels and Oliveri, Roberto and Orselli, Marta and Placidi, Andrea and Santos, Jorge E.",
    title = "{Blandford-Znajek monopole expansion revisited: novel non-analytic contributions to the power emission}",
    doi = "10.1088/1475-7516/2022/07/032",
    journal = "JCAP",
    volume = "07",
    number = "07",
    pages = "032",
    year = "2022"
}

@article{Tchekhovskoy:2009ba,
    author = "Tchekhovskoy, Alexander and Narayan, Ramesh and McKinney, Jonathan C.",
    title = "{Black Hole Spin and the Radio Loud/Quiet Dichotomy of Active Galactic Nuclei}",
    doi = "10.1088/0004-637X/711/1/50",
    journal = "Astrophys. J.",
    volume = "711",
    pages = "50--63",
    year = "2010"
}

@article{Meringolo:2025bdu,
    author = "Meringolo, Claudio and Camilloni, Filippo and Rezzolla, Luciano",
    title = "{Electromagnetic Energy Extraction from Kerr Black Holes: Ab Initio Calculations}",
    doi = "10.3847/2041-8213/ae06a6",
    journal = "Astrophys. J. Lett.",
    volume = "992",
    number = "1",
    pages = "L8",
    year = "2025"
}

@article{Camilloni:2023wyn,
    author = "Camilloni, Filippo and Harmark, Troels and Orselli, Marta and Rodriguez, Maria J.",
    title = "{Blandford-Znajek jets in MOdified Gravity}",
    doi = "10.1088/1475-7516/2024/01/047",
    journal = "JCAP",
    volume = "01",
    pages = "047",
    year = "2024"
}

@article{Randall:1999vf,
    author = "Randall, Lisa and Sundrum, Raman",
    title = "{An Alternative to compactification}",
    doi = "10.1103/PhysRevLett.83.4690",
    journal = "Phys. Rev. Lett.",
    volume = "83",
    pages = "4690--4693",
    year = "1999"
}

@article{Aliev:2005bi,
    author = "{A. N. Aliev and A. E. G\"{u}mr\"{u}k\c{c}\"{u}o\u{g}lu}",
    title = "{Charged rotating black holes on a 3-brane}",
    doi = "10.1103/PhysRevD.71.104027",
    journal = "Phys. Rev. D",
    volume = "71",
    pages = "104027",
    year = "2005"
}

@article{Pun:2008ua,
    author = "Pun, C. S. J. and Kovacs, Z. and Harko, T.",
    title = "{Thin accretion disks onto brane world black holes}",
    doi = "10.1103/PhysRevD.78.084015",
    journal = "Phys. Rev. D",
    volume = "78",
    pages = "084015",
    year = "2008"
}

@article{Schee:2008kz,
    author = "Schee, Jan and Stuchlik, Zdenek",
    title = "{Optical phenomena in the field of braneworld Kerr black holes}",
    doi = "10.1142/S0218271809014881",
    journal = "Int. J. Mod. Phys. D",
    volume = "18",
    pages = "983--1024",
    year = "2009"
}

@article{Stuchlik:2008fy,
    author = "Stuchl{\'\i}k, Zden{\v{e}}k and Kotrlov{\'a}, Andrea",
    title = "{Orbital resonances in discs around braneworld Kerr black holes}",
    doi = "10.1007/s10714-008-0709-2",
    journal = "Gen. Rel. Grav.",
    volume = "41",
    pages = "1305--1343",
    year = "2009"
}

@article{Abdujabbarov:2009az,
    author = "Abdujabbarov, Ahmadjon and Ahmedov, Bobomurat",
    title = "{Charged Particle Motion Around Rotating Black Hole in Braneworld Immersed in Magnetic Field}",
    doi = "10.1103/PhysRevD.81.044022",
    journal = "Phys. Rev. D",
    volume = "81",
    pages = "044022",
    year = "2010"
}

@article{Aliev:2009cg,
    author = "Aliev, Alikram N. and Talazan, Pamir",
    title = "{Gravitational Effects of Rotating Braneworld Black Holes}",
    doi = "10.1103/PhysRevD.80.044023",
    journal = "Phys. Rev. D",
    volume = "80",
    pages = "044023",
    year = "2009"
}

@article{Amarilla:2011fx,
    author = "Amarilla, Leonardo and Eiroa, Ernesto F.",
    title = "{Shadow of a rotating braneworld black hole}",
    doi = "10.1103/PhysRevD.85.064019",
    journal = "Phys. Rev. D",
    volume = "85",
    pages = "064019",
    year = "2012"
}

@article{Aliev:2013jqz,
    author = {Aliev, Alikram N. and Esmer, G{\"o}ksel Daylan and Talazan, Pamir},
    title = "{Strong Gravity Effects of Rotating Black Holes: Quasiperiodic Oscillations}",
    doi = "10.1088/0264-9381/30/4/045010",
    journal = "Class. Quant. Grav.",
    volume = "30",
    pages = "045010",
    year = "2013"
}

@article{Blaschke:2016uyo,
    author = "Blaschke, Martin and Stuchl{\'\i}k, Zden{\v{e}}k",
    title = "{Efficiency of the Keplerian accretion in braneworld Kerr-Newman spacetimes and mining instability of some naked singularity spacetimes}",
    doi = "10.1103/PhysRevD.94.086006",
    journal = "Phys. Rev. D",
    volume = "94",
    number = "8",
    pages = "086006",
    year = "2016"
}

@article{Stuchlik:2017rir,
    author = "Stuchl{\'\i}k, Zden{\v{e}}k and Blaschke, Martin and Schee, Jan",
    title = "{Particle collisions and optical effects in the mining Kerr-Newman spacetimes}",
    doi = "10.1103/PhysRevD.96.104050",
    journal = "Phys. Rev. D",
    volume = "96",
    number = "10",
    pages = "104050",
    year = "2017"
}

@article{Vagnozzi:2019apd,
    author = "Vagnozzi, Sunny and Visinelli, Luca",
    title = "{Hunting for extra dimensions in the shadow of M87*}",
    doi = "10.1103/PhysRevD.100.024020",
    journal = "Phys. Rev. D",
    volume = "100",
    number = "2",
    pages = "024020",
    year = "2019"
}

@article{Khan:2019gco,
    author = "Khan, Saeed Ullah and Shahzadi, Misbah and Ren, Jingli",
    title = "{Particle collisions in ergoregion of braneworld Kerr black hole}",
    doi = "10.1016/j.dark.2019.100331",
    journal = "Phys. Dark Univ.",
    volume = "26",
    pages = "100331",
    year = "2019"
}

@article{Banerjee:2019nnj,
    author = "Banerjee, Indrani and Chakraborty, Sumanta and SenGupta, Soumitra",
    title = "{Silhouette of M87*: A New Window to Peek into the World of Hidden Dimensions}",
    doi = "10.1103/PhysRevD.101.041301",
    journal = "Phys. Rev. D",
    volume = "101",
    number = "4",
    pages = "041301",
    year = "2020"
}

@article{Nucamendi:2019qsn,
    author = "Nucamendi, Ulises and Becerril, Ricardo and Sheoran, Pankaj",
    title = "{Bounds on spinning particles in their innermost stable circular orbits around rotating braneworld black hole}",
    doi = "10.1140/epjc/s10052-019-7584-8",
    journal = "Eur. Phys. J. C",
    volume = "80",
    number = "1",
    pages = "35",
    year = "2020"
}

@article{deOliveira:2020lzp,
    author = "de Oliveira, Ednilton S.",
    title = "{Tidal-charge effects on the superradiance of rotating black holes}",
    doi = "10.1140/epjc/s10052-020-08570-y",
    journal = "Eur. Phys. J. C",
    volume = "80",
    number = "11",
    pages = "1048",
    year = "2020"
}

@article{Neves:2020doc,
    author = "Neves, Juliano C. S.",
    title = "{Constraining the tidal charge of brane black holes using their shadows}",
    doi = "10.1140/epjc/s10052-020-8321-z",
    journal = "Eur. Phys. J. C",
    volume = "80",
    number = "8",
    pages = "717",
    year = "2020"
}

@article{deOliveira:2020jha,
    author = "de Oliveira, Ednilton S.",
    title = "{Scalar absorption cross section of rotating black holes with tidal charge}",
    doi = "10.1103/PhysRevD.104.124008",
    journal = "Phys. Rev. D",
    volume = "104",
    number = "12",
    pages = "124008",
    year = "2021"
}

@article{Dey:2020pth,
    author = "Dey, Ramit and Biswas, Shauvik and Chakraborty, Sumanta",
    title = "{Ergoregion instability and echoes for braneworld black holes: Scalar, electromagnetic, and gravitational perturbations}",
    doi = "10.1103/PhysRevD.103.084019",
    journal = "Phys. Rev. D",
    volume = "103",
    number = "8",
    pages = "084019",
    year = "2021"
}

@article{Hou:2021okc,
    author = "Hou, Yehui and Guo, Minyong and Chen, Bin",
    title = "{Revisiting the shadow of braneworld black holes}",
    doi = "10.1103/PhysRevD.104.024001",
    journal = "Phys. Rev. D",
    volume = "104",
    number = "2",
    pages = "024001",
    year = "2021"
}

@article{Chakraborty:2021gdf,
    author = "Chakraborty, Sumanta and Datta, Sayak and Sau, Subhadip",
    title = "{Tidal heating of black holes and exotic compact objects on the brane}",
    doi = "10.1103/PhysRevD.104.104001",
    journal = "Phys. Rev. D",
    volume = "104",
    number = "10",
    pages = "104001",
    year = "2021"
}

@article{Banerjee:2021aln,
    author = "Banerjee, Indrani and Chakraborty, Sumanta and SenGupta, Soumitra",
    title = "{Looking for extra dimensions in the observed quasi-periodic oscillations of black holes}",
    doi = "10.1088/1475-7516/2021/09/037",
    journal = "JCAP",
    volume = "09",
    pages = "037",
    year = "2021"
}

@article{Mishra:2021waw,
    author = "Mishra, Akash K. and Ghosh, Abhirup and Chakraborty, Sumanta",
    title = "{Constraining extra dimensions using observations of black hole quasi-normal modes}",
    doi = "10.1140/epjc/s10052-022-10788-x",
    journal = "Eur. Phys. J. C",
    volume = "82",
    number = "9",
    pages = "820",
    year = "2022"
}

@article{Biswas:2021gvq,
    author = "Biswas, Shauvik",
    title = "{Massive scalar perturbation of extremal rotating braneworld black hole: Superradiant stability analysis}",
    doi = "10.1016/j.physletb.2021.136597",
    journal = "Phys. Lett. B",
    volume = "820",
    pages = "136597",
    year = "2021"
}

@article{Du:2021czy,
    author = "Du, Yongbin and Liu, Yunlong and Zhang, Xiangdong",
    title = "{Collisional Penrose process of braneworld black hole with spinning particles}",
    doi = "10.1140/epjc/s10052-022-10833-9",
    journal = "Eur. Phys. J. C",
    volume = "82",
    number = "10",
    pages = "871",
    year = "2022"
}

@article{Wei:2022jbi,
    author = "Wei, Shao-Wen and Wang, Hui-Min and Zhang, Yu-Peng and Liu, Yu-Xiao",
    title = "{Effects of tidal charge on magnetic reconnection and energy extraction from spinning braneworld black hole}",
    doi = "10.1088/1475-7516/2022/04/050",
    journal = "JCAP",
    volume = "04",
    number = "04",
    pages = "050",
    year = "2022"
}

@article{Banerjee:2022jog,
    author = "Banerjee, Indrani and Chakraborty, Sumanta and SenGupta, Soumitra",
    title = "{Hunting extra dimensions in the shadow of Sgr A*}",
    doi = "10.1103/PhysRevD.106.084051",
    journal = "Phys. Rev. D",
    volume = "106",
    number = "8",
    pages = "084051",
    year = "2022"
}

@article{Bohra:2023vls,
    author = "Bohra, Sunil Singh and Sarkar, Subhodeep and Sen, Anjan Ananda",
    title = "{Gravitational atoms in the braneworld scenario}",
    doi = "10.1103/PhysRevD.109.104021",
    journal = "Phys. Rev. D",
    volume = "109",
    number = "10",
    pages = "104021",
    year = "2024"
}

@article{Zi:2024dpi,
    author = "Zi, Tieguang",
    title = "{Extreme mass-ratio inspiral as a probe of extra dimensions: The case of spinning massive object}",
    doi = "10.1016/j.physletb.2024.138538",
    journal = "Phys. Lett. B",
    volume = "850",
    pages = "138538",
    year = "2024"
}

@article{Liu:2024brf,
    author = "Liu, Ailin and He, Tong-Yu and Liu, Ming and Han, Zhan-Wen and Yang, Rong-Jia",
    title = "{Possible signatures of higher dimension in thin accretion disk around brane world black hole}",
    doi = "10.1088/1475-7516/2024/07/062",
    journal = "JCAP",
    volume = "07",
    pages = "062",
    year = "2024"
}

@article{Kumar:2025njz,
    author = "Kumar, Shailesh and Zi, Tieguang and Bhattacharyya, Arpan",
    title = "{Imprints of extra dimensions in eccentric extreme mass-ratio inspiral gravitational waveforms}",
    doi = "10.1103/xnxb-cjpl",
    journal = "Phys. Rev. D",
    volume = "112",
    number = "2",
    pages = "024039",
    year = "2025"
}

@article{Wang:2025fuw,
    author = "Wang, Hui-Min and Liao, Kai and Wei, Shao-Wen",
    title = "{Characteristic precessions of spherical orbit around a rotating braneworld black hole}",
    doi = "10.1140/epjc/s10052-025-14626-8",
    journal = "Eur. Phys. J. C",
    volume = "85",
    number = "9",
    pages = "933",
    year = "2025"
}

@article{Chamblin:2000ra,
    author = "Chamblin, Andrew and Reall, Harvey S. and Shinkai, Hisa-aki and Shiromizu, Tetsuya",
    title = "{Charged brane world black holes}",
    doi = "10.1103/PhysRevD.63.064015",
    journal = "Phys. Rev. D",
    volume = "63",
    pages = "064015",
    year = "2001"
}

@article{Dadhich:2000am,
    author = "Dadhich, Naresh and Maartens, Roy and Papadopoulos, Philippos and Rezania, Vahid",
    title = "{Black holes on the brane}",
    doi = "10.1016/S0370-2693(00)00798-X",
    journal = "Phys. Lett. B",
    volume = "487",
    pages = "1--6",
    year = "2000"
}

@article{Gralla:2014yja,
    author = "Gralla, Samuel E. and Jacobson, Ted",
    title = "{Spacetime approach to force-free magnetospheres}",
    doi = "10.1093/mnras/stu1690",
    journal = "Mon. Not. Roy. Astron. Soc.",
    volume = "445",
    number = "3",
    pages = "2500--2534",
    year = "2014"
}

@article{Znajek:1977,
    author = {Znajek, R. L.},
    title = {Black hole electrodynamics and the Carter tetrad},
    journal = {Mon. Not. Roy. Astron. Soc.},
    volume = {179},
    number = {3},
    pages = {457-472},
    year = {1977},
    month = {07},
    doi = {10.1093/mnras/179.3.457}
}

@article{Komissarov:2004ms,
    author = "Komissarov, S. S.",
    title = "{Electrodynamics of black hole magnetospheres}",
    doi = "10.1111/j.1365-2966.2004.07598.x",
    journal = "Mon. Not. Roy. Astron. Soc.",
    volume = "350",
    pages = "407",
    year = "2004"
}

@article{Nathanail:2014aua,
    author = "Nathanail, Antonios and Contopoulos, Ioannis",
    title = "{Black Hole Magnetospheres}",
    doi = "10.1088/0004-637X/788/2/186",
    journal = "Astrophys. J.",
    volume = "788",
    number = "2",
    pages = "186",
    year = "2014"
}

@article{Armas:2020mio,
    author = "Armas, Jay and Cai, Yangyang and Comp{\`e}re, Geoffrey and Garfinkle, David and Gralla, Samuel E.",
    title = "{Consistent Blandford-Znajek Expansion}",
    doi = "10.1088/1475-7516/2020/04/009",
    journal = "JCAP",
    volume = "04",
    pages = "009",
    year = "2020"
}

@article{Gralla:2015vta,
    author = "Gralla, Samuel E. and Lupsasca, Alexandru and Rodriguez, Maria J.",
    title = "{Electromagnetic Jets from Stars and Black Holes}",
    doi = "10.1103/PhysRevD.93.044038",
    journal = "Phys. Rev. D",
    volume = "93",
    number = "4",
    pages = "044038",
    year = "2016"
}

@article{Gralla:2015uta,
    author = "Gralla, Samuel E. and Lupsasca, Alexandru and Rodriguez, Maria J.",
    title = "{Note on Bunching of Field Lines in Black Hole Magnetospheres}",
    doi = "10.1103/PhysRevD.92.044053",
    journal = "Phys. Rev. D",
    volume = "92",
    number = "4",
    pages = "044053",
    year = "2015"
}

@article{Tanabe:2008wm,
    author = "Tanabe, Kentarou and Nagataki, Shigehiro",
    title = "{Extended monopole solution of the Blandford-Znajek mechanism: Higher order terms for a Kerr parameter}",
    doi = "10.1103/PhysRevD.78.024004",
    journal = "Phys. Rev. D",
    volume = "78",
    pages = "024004",
    year = "2008"
}

@article{Konoplya:2021qll,
    author = "Konoplya, R. A. and Kunz, J. and Zhidenko, A.",
    title = "{Blandford-Znajek mechanism in the general stationary axially-symmetric black-hole spacetime}",
    doi = "10.1088/1475-7516/2021/12/002",
    journal = "JCAP",
    volume = "12",
    number = "12",
    pages = "002",
    year = "2021"
}

@article{Krishnendu:2024jkj,
    author = "Krishnendu, N. V. and Chakraborty, Sumanta",
    title = "{Probing black hole charge from the binary black hole inspiral}",
    doi = "10.1103/PhysRevD.109.124047",
    journal = "Phys. Rev. D",
    volume = "109",
    number = "12",
    pages = "124047",
    year = "2024"
}

@article{Figueras:2011gd,
    author = "Figueras, Pau and Wiseman, Toby",
    title = "{Gravity and large black holes in Randall-Sundrum II braneworlds}",
    doi = "10.1103/PhysRevLett.107.081101",
    journal = "Phys. Rev. Lett.",
    volume = "107",
    pages = "081101",
    year = "2011"
}

@article{Biggs:2021iqw,
    author = "Biggs, William D. and Santos, Jorge E.",
    title = "{Rotating Black Holes in Randall-Sundrum II Braneworlds}",
    doi = "10.1103/PhysRevLett.128.021601",
    journal = "Phys. Rev. Lett.",
    volume = "128",
    number = "2",
    pages = "021601",
    year = "2022"
}
\bibliographystyle{sn-aps}
%%%%%%%%%%%%%%%%%%%%%%%%%%%%%%%%%%%%%%%%%%%%%%%%%%%%%%%%%%%%%%%%%%%%%%%%%%%%
\end{document}